\documentclass[aps,prl,twocolumn,groupedaddress]{revtex4-1}

\usepackage{graphicx}
\usepackage{bm}
\usepackage{tabularx}
\usepackage{amsmath}
\usepackage{latexsym} 

\begin{document}

\preprint{}

\title{ Planar CuO$_2$ hole density estimation in multilayered high-$T_c$ cuprates 
}

\author{Sunao Shimizu}
\email[]{e-mail: shimizu@nmr.mp.es.osaka-u.ac.jp}
\author{Shiho Iwai} 
\author{Shin-ichiro Tabata}
\author{Hidekazu Mukuda}
\author{Yoshio Kitaoka}
\affiliation{Department of Materials Engineering Science, Osaka University, Osaka 560-8531, Japan }
\author{Parasharam M. Shirage}
\author{Hijiri Kito}
\author{Akira Iyo}
\affiliation{National Institute of Advanced Industrial Science and Technology (AIST), Umezono, Tsukuba 305-8568, Japan}

\date{\today}

\begin{abstract}
We report that planar CuO$_2$ hole densities in high-$T_c$ cuprates are consistently determined by the Cu-NMR Knight shift. In single- and bi-layered cuprates, it is demonstrated that the spin part of the Knight shift $K_s$(300 K) at room temperature monotonically increases with the hole density $p$ from underdoped to overdoped regions, suggesting that the relationship of $K_s$(300 K) vs. $p$ is a reliable measure to determine $p$. 
The validity of this $K_s$(300 K)-$p$ relationship is confirmed by the investigation of the $p$-dependencies of hyperfine magnetic fields and of spin susceptibility for single- and bi-layered cuprates with tetragonal symmetry.
Moreover, the analyses are compared with the NMR data on three-layered Ba$_2$Ca$_2$Cu$_3$O$_6$(F,O)$_2$, HgBa$_2$Ca$_2$Cu$_3$O$_{8+\delta}$, and five-layered HgBa$_2$Ca$_4$Cu$_5$O$_{12+\delta}$, which suggests the general applicability of the $K_s$(300 K)-$p$ relationship to multilayered compounds with more than three CuO$_2$ planes. We remark that the measurement of $K_s$(300 K) enables us to separately estimate $p$ for each CuO$_2$ plane in multilayered compounds, where doped hole carriers are inequivalent between outer CuO$_2$ planes and inner CuO$_2$ planes.
\end{abstract}

\pacs{74.72.Jt; 74.25.Ha; 74.25.Nf}

\maketitle

\section{Introduction}
 
In hole doped high-$T_c$ cuprates, the hole density $p$ in CuO$_2$ planes has been determined by various methods: indirect chemical methods like solid solutions \cite{Shafer,Torrance,Kishino,Tokura}, bond valence sums determined from structural bond lengths \cite{Brown,TallonBV,Cava}, or the Fermi surface topology \cite{Kordyuk}. Moreover, the thermoelectric power is a universal function of $p$ \cite{Obertelli,Tallon}, and the phase diagram for hole doped cuprates is well described by $T_c$=$T_{c,max}[1-82.6(p-0.16)^2]$ \cite{Presland}, which is applicable for the estimation of $p$ when suitable structural data are not available.  
These methods are, however, inapplicable when we deal with multilayered compounds because they are composed of more than two inequivalent CuO$_2$ planes in a unit cell (see Fig.~\ref{fig:n3NMR}(a) as an example). In multilayered cuprates, doped hole carriers are inequivalent between outer CuO$_2$ planes (OPs) and inner CuO$_2$ planes (IPs) due to the imbalance of the Madelung potential on each CuO$_2$ plane; the above-mentioned methods would evaluate not {\it each hole density} inherent to CuO$_2$ planes but a {\it total hole density}.
Precise determination of hole density on respective CuO$_2$ planes is indispensable for gaining deep insight into the phenomena observed in multilayered cuprates \cite{Tokunaga,Kotegawa2001,Mukuda2008,ShimizuJPSJ,ShimizuFIN}.

It is known that the spin part of the Knight shift $K_s$(300 K) at room temperature increases with $p$ in hole doped cuprates \cite{Kotegawa2001,Walstedt,Ohsugi,Ishida,FujiwaraJPSJ,MagishiTl1212,Storey}. 
The $K_s$(300 K) for OP and IP are separately determined by Cu-NMR. Therefore, each value of $p$ for OP and for IP can be estimated if the relationship between $K_s$(300 K) and $p$ is established.
Meanwhile, a linear equation $p$=0.502$K_s$(300 K)+0.0462 has been reported \cite{Kotegawa2001,TokunagaJLTP}, where $p$ was derived from the NQR frequencies of Cu and O in CuO$_2$ planes \cite{Zheng}.  
The values of $p$ estimated by the linear equation, however, seem relatively large. For example, an optimal doping level has been evaluated to be $p$ $\sim$ 0.22-0.24 in five-layered Hg-based compounds \cite{Mukuda2008}, the value of which is larger than a widely-accepted optimal doping level in hole-doped cuprates, $p$ $\sim$ 0.16. This inconsistency is probably, in part, due to the calculation that connects NQR frequencies to $p$ \cite{Haase}. 
In addition to that, it is nontrivial whether the $p$-dependence of $K_s$(300 K) is described with a single equation which holds among different materials. $K_s$ is proportional to spin susceptibility $\chi_s$ and hyperfine coupling constants $A_{hf}$ as $K_s$=$A_{hf} \chi_s$, where the supertransferred magnetic field $B$ in $A_{hf}$=$A$+4$B$ depends on the materials \cite{MR}.

In this paper, we report that the planar CuO$_2$ hole density $p$ in high-$T_c$ cuprates is consistently evaluated from the spin part of the $^{63}$Cu-Knight shift $K_s$(300 K) at room temperature.
We show the $p$ dependences of $K_s$(300K) and of $B$ in bi-layered Ba$_2$CaCu$_2$O$_4$(F,O)$_2$ (0212F), together with other single- and bi-layered materials with tetragonal symmetry.
NMR data on three-layered Ba$_2$Ca$_2$Cu$_3$O$_6$(F,O)$_2$ (0223F), HgBa$_2$Ca$_2$Cu$_3$O$_{8+\delta}$ (Hg1223) \cite{MagishiHg1223}, and HgBa$_2$Ca$_4$Cu$_5$O$_{12+\delta}$ (Hg1245) \cite{Kotegawa2004} suggest that the $p$-dependencies of $K_s$(300 K) and of $B$ hold in multilayered compounds as well.
These results show that the present method, based on the Knight shift, enables us to separately estimate $p$ for each CuO$_2$ plane in multilayered high-$T_c$ cuprates.

\begin{table*}[htbp]
\caption[]{List of samples measured in this study, bi-layered Ba$_2$CaCu$_2$O$_4$(F,O)$_2$ (0212F) and three-layered Ba$_2$Ca$_2$Cu$_3$O$_6$(F,O)$_2$ (0223F). $T_c$ was determined by the onset of SC diamagnetism using a dc SQUID magnetometer. The values of $p$ for 0212F are evaluated using $T_c$=$T_{c,max}[1-82.6(p-0.16)^2]$ \cite{Presland}. For three-layered 0223F, two values of $K_{s,ab}$(300 K) and of $B$ correspond to OP/IP.}
\label{samples}
\begin{center}
{\tabcolsep = 3.5mm
 \renewcommand\arraystretch{1.2}
  \begin{tabular}{cccccc}
    \hline\hline
Sample           &$T_c$(K)&    $p$      &  $K_{s,ab}$(300 K) ($\%$) & $B$(kOe)     \\
    \hline
0212F($\sharp$1) &    102 &      0.174  &      0.38       &   80                   \\
0212F($\sharp$2) &    105 &      0.149  &      0.33       &   74                   \\
0212F($\sharp$3) &     73 &      0.114  &      0.25       &   67                   \\
0212F($\sharp$4) &    40  &      0.085  &      0.20       &   67                   \\
    \hline
0223F            &    120 &             &   0.349/0.275   &   72/67                \\

    \hline\hline
    \end{tabular}}
 \end{center}
\label{t:ggg}
\end{table*}

\section{Experimental details}

Polycrystalline powder samples of Ba$_2$CaCu$_2$O$_4$(F,O)$_2$ (0212F) and Ba$_2$Ca$_2$Cu$_3$O$_6$(F,O)$_2$ (0223F), which are listed in Table \ref{t:ggg}, were prepared by high-pressure synthesis, as described elsewhere \cite{Shirage,Iyo1,Iyo2}. The crystal structures are shown in Fig.~\ref{fig:n2NMR}(a) and in Fig.~\ref{fig:n3NMR}(a).
Powder X-ray diffraction analysis shows that these compounds comprise almost a single phase, and that the $a$-axis length continually decreases with an increase in the nominal fraction of O$^{2-}$ \cite{Shirage}.
$T_c$ was uniquely determined by the onset of SC diamagnetism using a dc SQUID magnetometer. 
Four 0212F samples exhibit a systematic change in $T_c$, as the nominal fraction of oxygen O$^{2-}$ decreases at the apical fluorine F$^{1-}$ sites, i.e., the hole doping level decreases. Note that the actual fraction of F$^{1-}$ and O$^{2-}$ is difficult to determine \cite{Shirage,ShimizuPRB,ShimizuPRL}. In Table \ref{t:ggg}, the hole density $p$ is evaluated by using $T_c$=$T_{c,max}[1-82.6(p-0.16)^2]$ \cite{Presland}.
For NMR measurements, the powder samples were aligned along the $c$-axis in an external field $H$ of $\sim$ 16 T and fixed using stycast 1266 epoxy. The NMR experiments were performed by a conventional spin-echo method in the temperature ($T$) range of 4.2 $-$ 300 K with $H$ perpendicular or parallel to the $c$-axis. The width of the first exciting $\pi$/2-pulse was 6 $\mu$s. The $H$ was calibrated by using the $^{27}$Al Free Induction Decay signal.
 
\section{Results}

\subsection{Bi-layered Ba$_2$CaCu$_2$O$_4$(F,O)$_2$}

\subsubsection{$^{63}$Cu-NMR with $H\parallel ab$}

\begin{figure}[htpb]
\begin{center}
\includegraphics[width=1.0\linewidth]{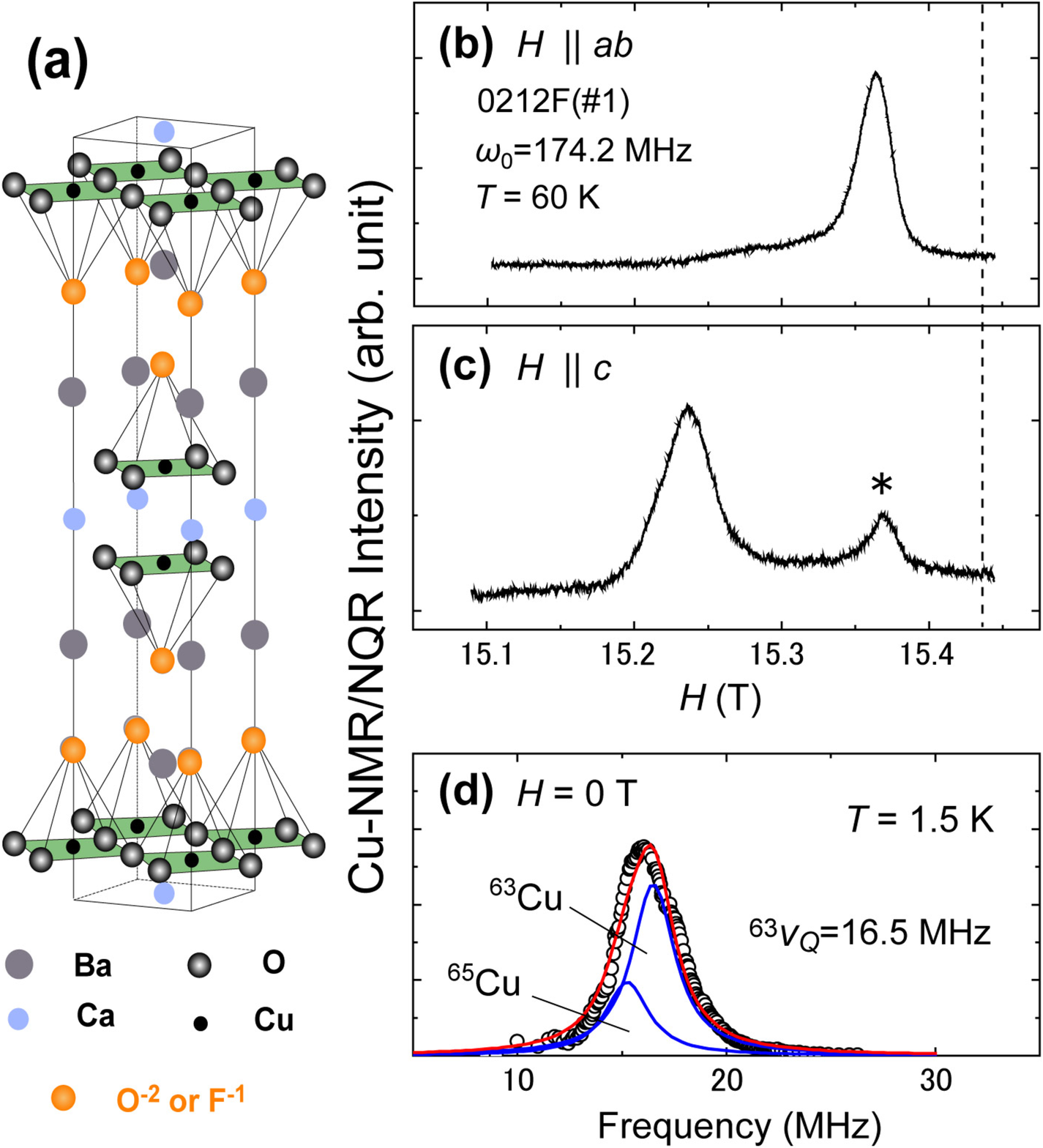}
\end{center}
\caption{\footnotesize (color online)  (a) Crystal structure of bi-layered Ba$_2$CaCu$_2$O$_4$(F,O)$_2$ (0212F). The heterovalent substitution of O$^{2-}$ for F$^{1-}$ increases the hole density. (b) $^{63}$Cu-NMR spectra for 0212F($\sharp$1) at $T$=60 K with $H$ perpendicular to the $c$-axis ($H\parallel ab$). The dashed line points to $K$=0. (c) $^{63}$Cu-NMR spectra for 0212F($\sharp$1) at $T$=60 K with $H$ parallel to the $c$-axis ($H\parallel c$). The peak marked with an asterisk ($\ast$) is from unoriented grains with $\theta$ $\sim$ 90$^{\circ}$, where $\theta$ is the angle between $H$ and the $c$-axis. 
If $\theta$ in the unoriented grains were under the completely random distribution, the peak ($\ast$) could coincide with the spectral peak in (b).  (d) Cu-NQR spectrum at $T$=1.5 K. The spectrum is composed of $^{63}$Cu and $^{65}$Cu.}
\label{fig:n2NMR}
\end{figure}

Figure~\ref{fig:n2NMR}(b) shows a typical $^{63}$Cu-NMR spectrum of the central transition (1/2 $\Leftrightarrow$ $-$1/2) for 0212F($\sharp$1), which has the largest nominal O$^{2-}$ composition among the bi-layered samples used in this study. The field-swept NMR spectrum was measured at $H$ perpendicular to the $c$-axis ($H\parallel ab$). Here, the NMR frequency $\omega_0$ was fixed at 174.2 MHz. 
According to the second order perturbation theory for the nuclear Hamiltonian \cite{Abragam,TakigawaNQRshift}, total NMR shifts consist of the Knight shift $K_{ab}$ with $H\parallel ab$ and the second order quadrupole shift as 
\begin{equation}
\frac{\omega_0 - \gamma_N H_{res}}{\gamma_N H_{res}} = K_{ab}+\frac{3\nu_Q^2}{16(1+K_{ab})}\frac{1}{(\gamma_N H_{res})^2}~, 
\label{eq:shift}
\end{equation}
where $\gamma_N$ is a nuclear gyromagnetic ratio, $H_{res}$ an NMR resonance field, and $\nu_Q$ the nuclear quadrupole frequency. In order to subtract $K_{ab}$ from the total NMR shift, we estimated the $\nu_Q$ by nuclear quadrupole resonance (NQR) measurements at $H$=0 T and $T$=1.5 K. The NQR spectrum for 0212F($\sharp$1) is shown in Fig.~\ref{fig:n2NMR}(d), which has the $^{63}$Cu and $^{65}$Cu components. The $^{63}\nu_Q$ of $^{63}$Cu is estimated to be 16.5 MHz by the spectral fitting shown in the figure. For other bi-layered compounds, the values of $^{63}\nu_Q$ are 15.7, 13.7, and 12.5 MHz for 0212F($\sharp$2), ($\sharp$3), and ($\sharp$4), respectively.

\begin{figure}[tpb]
\begin{center}
\includegraphics[width=0.8\linewidth]{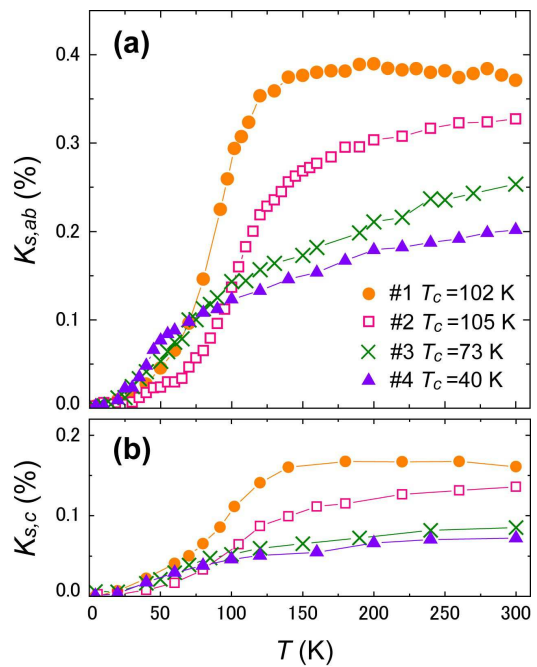}
\end{center}
\caption{\footnotesize (color online)  (a) Spin part of the $^{63}$Cu Knight shift $K_{s,ab}(T)$ with $H\parallel ab$ as a function of temperature. The data plots are assigned as labeled in the figure. (b) Spin part of the $^{63}$Cu Knight shift $K_{s,c}(T)$ with $H\parallel c$. $K_{s,c}(T)$ shows the similar $T$-dependence with $K_{s,ab}(T)$.}
\label{fig:n2KS}
\end{figure}

Figure \ref{fig:n2KS}(a) shows the spin part of the Knight shift, $K_{s,ab}(T)$, as a function of temperature.
The Knight shift $K$ in high-$T_c$ cuprates comprises a $T$-dependent spin part $K_s(T)$ and a $T$-independent orbital part $K_{orb}$ as follows:
\begin{equation}
K_\alpha  = K_{s,\alpha}(T)+K_{orb,\alpha}  ~~~(\alpha =c,ab), 
\label{eq:K}
\end{equation}
where $\alpha$ is the direction of $H$.
We estimate $K_{orb,ab}$ $\simeq$ 0.23 $\%$, assuming $K_{s,ab}$ $\simeq$ 0 at $T$=0 limit. The value of $K_{orb,ab}$ is consistent with those in other hole-doped high-$T_c$ cuprates \cite{Walstedt,IshidaBi2212,FujiwaraJPSJ,MagishiTl1212}.
Upon cooling down to $T_c$, $K_{s,ab}(T)$ is roughly constant for 0212F($\sharp$1), while $K_{s,ab}(T)$ decreases for the other four samples in association with the opening of pseudogaps \cite{Yasuoka,REbook}. It is well known that the pseudogaps in $K_{s,ab}(T)$ emerge in underdoped regions, not in overdoped regions \cite{IshidaBi2212,FujiwaraJPSJ}. A steep decrease below $T_c$ for all samples indicates the reduction in spin susceptibility $\chi_s(T)$ proportional to $K_{s,ab}(T)$ due to the formation of spin-singlet Cooper pairing. 
We note here that in hole-doped cuprates $K_{s,ab}(T)$ at room temperature, $K_{s,ab}$(300 K), increases with hole density $p$ \cite{Kotegawa2001,Walstedt,Ohsugi,Ishida,FujiwaraJPSJ,MagishiTl1212,Storey}. The values of $K_{s,ab}$(300 K) in the 0212F samples are listed in Table.~\ref{t:ggg}. The relationship between $p$ and $K_{s,ab}$(300 K) is discussed later.

\subsubsection{$^{63}$Cu-NMR with $H\parallel c$}

Figure \ref{fig:n2NMR}(c) shows a typical Cu-NMR spectrum of the central transition (1/2 $\Leftrightarrow$ $-$1/2) for 0212F($\sharp$1), where the field-swept NMR spectra were measured at $H$ parallel to the $c$-axis ($H\parallel c$) and at $\omega_0$=174.2 MHz. The small peak denoted by the asterisk ($\ast$) in the higher-field region arises from unoriented grains with $\theta$ $\sim$ 90$^{\circ}$, where $\theta$ is the angle between $H$ and the $c$-axis. 

Figure \ref{fig:n2KS}(b) shows the $T$-dependence of $K_{s,c}(T)$. We estimate $K_{orb,c}$ $\simeq$ 1.22 $\%$, assuming $K_{s,c}$ $\simeq$ 0 at $T$=0 limit. Note that when $H\parallel c$-axis, the second order quadrupole shift, corresponding to the second term in Eq.~(\ref{eq:shift}), is zero \cite{Abragam,TakigawaNQRshift}. 
The $K_{s,c}(T)$ in Fig.~\ref{fig:n2KS}(b) shows a similar $T$-dependence with $K_{s,ab}(T)$ in Fig.~\ref{fig:n2KS}(a), as reported for other compounds such as Tl- \cite{MagishiTl1212,Kitaoka}, Bi- \cite{Ishida}, and Hg-based compounds \cite{Julien,MagishiHg1223}. It is, however, different from L214 \cite{Ohsugi}, YBa$_2$Cu$_3$O$_{6+x}$ (Y123) \cite{Walstedt,Barrett,Takigawa} and YBa$_2$Cu$_4$O$_8$ (Y124) \cite{Zimmermann}; $K_{s,c}(T)$ {\it increases} below $T_c$ upon cooling. This inconsistency comes from the difference of hyperfine magnetic fields discussed below.

\subsubsection{Hyperfine magnetic field in CuO$_2$ plane}

According to the Mila-Rice Hamiltonian \cite{MR}, the spin Knight shift of Cu in the CuO$_2$ plane is expressed as 
\begin{equation}
K_{s,\alpha}(T) = (A_{\alpha}+4B)\chi_s(T)  ~~~(\alpha =c,ab), 
\label{eq:Ks}
\end{equation}
where $A_{\alpha}$ and $B$ are the on-site and the supertransferred hyperfine fields of Cu, respectively. Here, the $A_{\alpha}$ consists of the contributions induced by on-site Cu $3d_{x^2-y^2}$ spins $-$ anisotropic dipole, spin-orbit, and isotropic core polarization; and the $B$ originates from the isotropic $4s$ spin polarization produced by neighboring four Cu spins through the Cu($3d_{x^2-y^2}$)-O($2p\sigma$)-Cu($4s$) hybridization. 
Since the spin susceptibility $\chi_s(T)$ is assumed to be isotropic, the anisotropy $\Delta$ of $K_{s,\alpha}(T)$ is given by 
\begin{equation}
\Delta \equiv \frac{K_{s,c}(T)}{K_{s,ab}(T)}=\frac{A_{c}+4B}{A_{ab}+4B}. 
\label{eq:B}
\end{equation}
From Fig.~\ref{fig:n2KS}, $\Delta$ is evaluated as $\sim$ 0.42, 0.38, 0.33, and  0.33 for 0212F($\sharp$1), 0212F($\sharp$2), 0212F($\sharp$3), and 0212F($\sharp$4), respectively. The on-site hyperfine fields, $A_{ab}$ $\sim$ 37 kOe/$\mu_B$ and $A_c$ $\sim$ $-$170 kOe/$\mu_B$ \cite{Monien,Millis,Imai}, are assumed as material-independent in hole-doped high-$T_c$ cuprates, which allows us to estimate $B$ for 0212F samples as listed in Table \ref{t:ggg}. 
These $B$ values are larger than $B\sim$ 40 kOe/$\mu_B$, which is a typical value for  L214 \cite{Ohsugi}, Y123 \cite{Walstedt,Barrett,Takigawa}, and Y124 \cite{Zimmermann} compounds.

\subsection{Three-layered Ba$_2$Ca$_2$Cu$_3$O$_6$(F,O)$_2$ }

\begin{figure}[htpb]
\begin{center}
\includegraphics[width=1.0\linewidth]{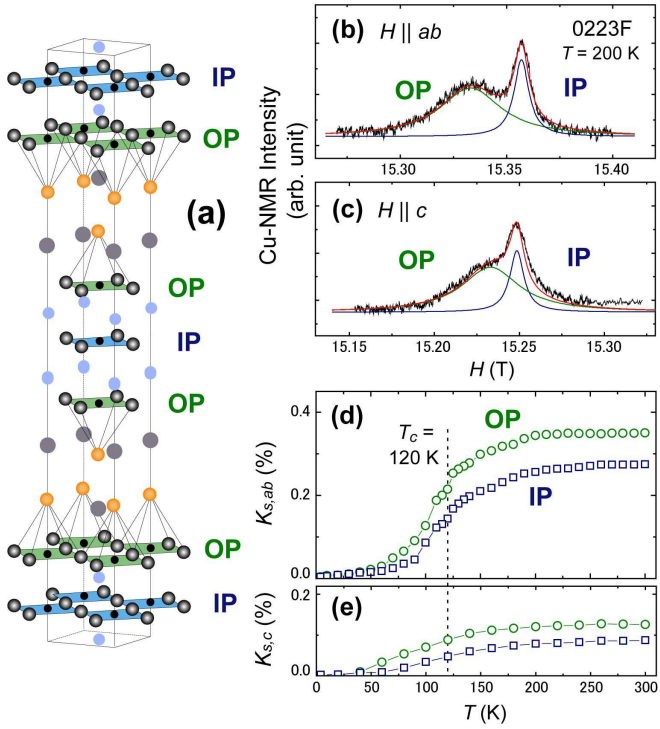}
\end{center}
\caption{\footnotesize (color online)  (a) Crystal structure of three-layered Ba$_2$Ca$_2$Cu$_3$O$_6$(F,O)$_2$ (0223F). In a unit cell, there are two kinds of CuO$_2$ planes: outer CuO$_2$ plane (OP) and inner CuO$_2$ plane (IP). (b,c) Cu-NMR spectrum of 0223F at $T$=200 K with $H\parallel ab$ and $H\parallel c$. The spectra have two components arising from OP and IP. (d,e) $T$-dependences of $K_{s,ab}(T)$ and $K_{s,c}(T)$ for 0223F.}
\label{fig:n3NMR}
\end{figure}

Figures \ref{fig:n3NMR}(b) and \ref{fig:n3NMR}(c) show the Cu-NMR spectra of the central transition (1/2 $\Leftrightarrow$ $-$1/2) for three-layered Ba$_2$Ca$_2$Cu$_3$O$_6$(F,O)$_2$ (0223F) at $T$=200 K. The field-swept NMR spectra in Figs.~\ref{fig:n3NMR}(a) and \ref{fig:n3NMR}(b) were measured at $H\parallel ab$ and $H\parallel c$, respectively. Here, the NMR frequency $\omega_0$ was fixed at 174.2 MHz. 
As shown in the crystal structure of 0223F in Fig.~\ref{fig:n3NMR}(a), multilayered compounds, which have more than three CuO$_2$ planes in a unit cell, are composed of two kinds of CuO$_2$ planes: outer CuO$_2$ plane (OP) and inner CuO$_2$ plane (IP).
The NMR spectra in Figs.~\ref{fig:n3NMR}(b) and \ref{fig:n3NMR}(c) were assigned to OP and IP as denoted in the figure according to the literature \cite{JulienPRL}.
The NMR spectral width for OP is much broader than that for IP; OP is closer to the Ba-F layer (see Fig.~\ref{fig:n3NMR}(a)), which is the source of the disorder due to the atomic substitution at apical-F sites.

Figures \ref{fig:n3NMR}(d) and \ref{fig:n3NMR}(e) show the $T$-dependences of $K_{s,ab}(T)$ and $K_{s,c}(T)$ for 0223F, respectively. Above $T_c$=120 K, both $K_{s,ab}$ and $K_{s,c}$ decrease upon cooling $T$ due to the opening of pseudogaps, which suggests that both OP and IP are underdoped. 
As mentioned before in connection with Fig.~\ref{fig:n2KS}(a), $K_{s,ab}$(300 K) increases with $p$. 
Therefore, Fig.~\ref{fig:n3NMR}(d) suggests that $p$(OP) is larger than $p$(IP) in 0223F. The estimation of hole density in multilayered cuprates is discussed later. The anisotropy $\Delta$ is evaluated from $K_{s,ab}(T)$ and $K_{s,c}(T)$ (see Eq.~(\ref{eq:B})), as conducted in bi-layered 0212F samples. $\Delta$(OP) and $\Delta$(IP) are evaluated as $\sim$ 0.36 and $\sim$ 0.32, which provide $B$(OP) $\sim$ 72 kOe/$\mu_B$ and $B$(IP) $\sim$ 67 kOe/$\mu_B$.

\section{Discussions}

\subsection{$p$ dependence of $K_{s,ab}$(300 K), $B$, and $\chi_s$(300 K) in plane Cu site}

\begin{figure}[htpb]
\begin{center}
\includegraphics[width=0.85\linewidth]{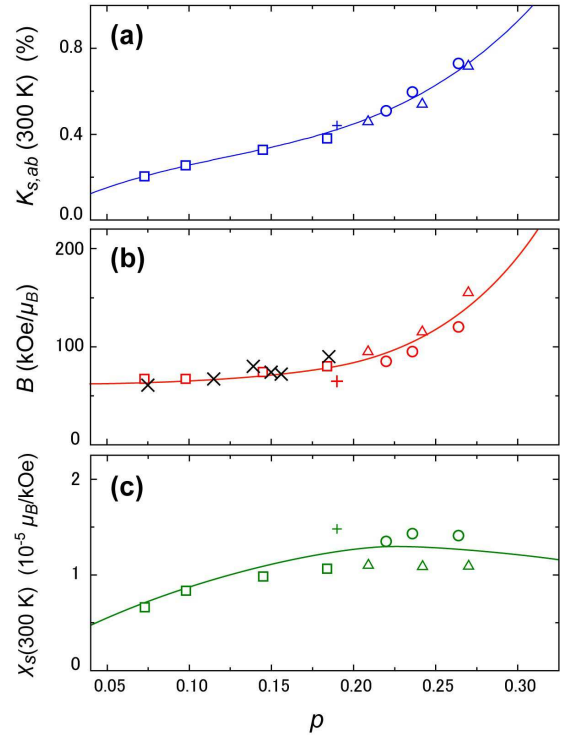}
\end{center}
\caption{\footnotesize (color online)  (a) $K_{s,ab}$(300 K) and (b) $B$ are plotted as a function of $p$ for 0212F ($\Box$), Bi2212 ($+$), Tl2201 ($\triangle$), and Tl1212 ($\bigcirc$). The values of $p$ are estimated from $T_c$=$T_{c,max}[1-82.6(p-0.16)^2]$ \cite{Presland}. The plots denoted by crosses ($\times$) in (b) are for multilayered compounds, which are listed in Table \ref{t:ggg2}. (c) $p$-dependence of $\chi_s$(300 K) evaluated from $K_{s,ab}$=($A_{ab}+4B$)$\chi_s$ (Eq.~(\ref{eq:Ks})). The solid lines in the figures are guides to the eye.}
\label{fig:Bterm}
\end{figure}

Figure \ref{fig:Bterm}(a) shows $K_{s,ab}$(300 K) for 0212F, Bi$_2$Sr$_2$CaCu$_2$O$_8$ (Bi2212) \cite{IshidaBi2212}, Tl$_2$Ba$_2$CuO$_{6+\delta}$ (Tl2201) \cite{KitaokaTl2201}, and TlSr$_2$CaCu$_2$O$_{7-\delta}$ (Tl1212) \cite{MagishiTl1212} plotted against the $p$ evaluated by using $T_c$=$T_{c,max}[1-82.6(p-0.16)^2]$ \cite{Presland}. 
We note that those compounds have one or two CuO$_2$ planes with tetragonal symmetry, which are homologous series of the apical-F, Tl-, and Hg-based multilayered compounds. 
As shown in Fig.~\ref{fig:Bterm}(a), $K_{s,ab}$(300 K) monotonically increases with $p$ from underdoped to overdoped regions. $K_{s,ab}$(300 K) seems material-independent, suggesting that $K_s$(300 K) is a good indication of $p$ in hole-doped cuprates. If the validity of the $K_{s,ab}$(300 K)-$p$ relationship is presented, we can apply it to the estimation of $p$ in multilayered compounds.

\begin{figure}[htpb]
\begin{center}
\includegraphics[width=0.9\linewidth]{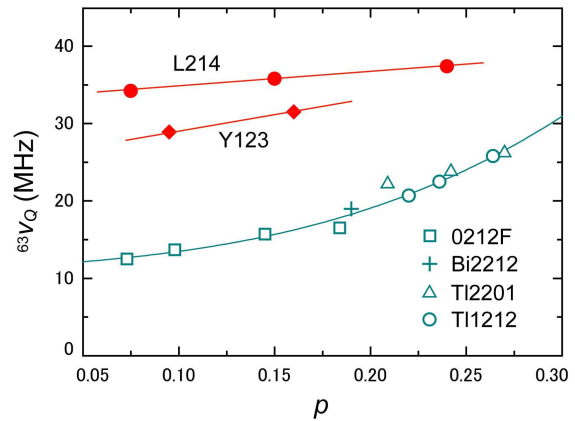}
\end{center}
\caption{\footnotesize (color online)  $^{63}$Cu-NQR frequency $^{63}\nu_Q$ for hole-doped cuprates, 0212F, Bi2212 \cite{IshidaBi2212}, Tl2201 \cite{Fujiwara}, Tl1212 \cite{MagishiTl1212}, L214 \cite{Ohsugi}, and Y123 \cite{Pennington,Yoshinari}. Here, $p$ is evaluated from Sr-content $x$ for L214 and from $T_{c,max}$ \cite{Presland} for Y123. The solid lines in the figure are guides to the eye.}
\label{fig:nuQ}
\end{figure}

According to Eq.~(\ref{eq:Ks}), the $p$ dependence of $K_{s,ab}$(300 K) is derived from those of $B$ and $\chi_s$(300 K). Figure \ref{fig:Bterm}(b) shows the $p$ dependence of $B$ for the same materials shown in Fig.~\ref{fig:Bterm}(a) \cite{IshidaBi2212,KitaokaTl2201,MagishiTl1212}. As presented in Fig.~\ref{fig:Bterm}(b), $B$ increases with $p$, showing a steep increase at $p$ $\sim$ 0.18-0.20. 
It is remarkable that the $B$ term exhibits weak $p$-dependence in the underdoped region, while it shows a steep increase with $p$ $>$ 0.16 in the overdoped region.
The $B$ term arises from the Cu($3d_{x^2-y^2}$)-O($2p\sigma$)-Cu($4s$) covalent bonds with the four nearest neighbor Cu sites; therefore, it is expected that the hybridization
between Cu($3d_{x^2-y^2}$) and O($2p\sigma$) orbits becomes larger as $p$ increases in an overdoped regime, and that as a result, $T_c$ starts to decrease.
In Fig.~\ref{fig:Bterm}(b), the $p$-dependent $B$ term seems {\it material-independent}; however, the $B$ terms for L214, Y123, and Y124, $\sim$ 40 kOe/$\mu_B$, are relatively small compared with the values shown in Fig.~\ref{fig:Bterm}(b).
This inconsistency is seen in the variation of $^{63}\nu_Q$ in Fig.~\ref{fig:nuQ} as well. The $^{63}\nu_Q$ increases with $p$ for all materials, while, for a given $p$, the absolute values for L214 and Y123 are about 2 to 3 times larger than those for others. The values of $^{63}\nu_Q$ depend on the hole number $n_d$ in Cu($3d_{x^2-y^2}$) orbit and $n_p$ in O($2p\sigma$) orbit. Therefore, it is expected that $n_d$ and $n_p$ in L214 and Y123 are distributed in a different manner from those in the cuprates presented here, even though 
$p$ ($p$=$n_d$+2$n_p-$1) is the same between the former and the latter.
Actually, it has been reported that $n_d$ is large in L214 and Y123, which is the reason for the large $\nu_Q$ \cite{Zheng,Haase}.  
In this context, it is considered that the hybridization between Cu($3d_{x^2-y^2}$) and O($2p\sigma$) orbits in L214,Y123, and Y124 is smaller than those in the cuprates presented here, resulting in the $B$ for the former being remarkably smaller than for the latter.
This is probably related to the crystal structures; L214, Y123, and Y124 have orthorhombic crystal structures in superconducting region, while 0212F, Bi2212, Tl2201, and Tl1212 have tetragonal ones. We conclude that the $p$-dependence of $B$ in Fig.~\ref{fig:Bterm}(b) holds in CuO$_2$ planes with tetragonal symmetry.

Figure \ref{fig:Bterm}(c) shows the $p$ dependence of $\chi_s$(300 K), evaluated using $K_{s,ab}$(300 K)=$|A_{ab}+4B|\chi_s$(300 K) (see Eq.~(\ref{eq:Ks})). 
The $\chi_s$(300 K) increases with $p$ and is nearly constant above $p$ $\sim$ 0.20-0.25, which is consistent with previous reports \cite{Loram}. 
In general, the spin susceptibility $\chi_s$ is related to the density of states at the Fermi level. Therefore, the reduction of $\chi_s$(300 K) in an underdoped regime would be attributed to the emergence of pseudogaps as discussed in previous literature \cite{Loram}. 

When we take into account the $p$-dependencies of $B$ and $\chi_s$(300 K), the $p$-dependence of $K_{s,ab}$(300 K) -- a monotonically increasing function -- is explained. 
In over-doped regimes, a strong hybridization between Cu($3d_{x^2-y^2}$) and O($2p\sigma$) orbits increases the $B$, making $K_{s,ab}$(300 K) larger as $p$ increases. In under-doped regimes, the opening of the pseudogap decreases $\chi_s$(300 K), making $K_{s,ab}$(300 K) smaller as $p$ decreases.
As a result, we conclude that the $p$-dependence of $K_{s,ab}$(300 K) holds in CuO$_2$ planes with tetragonal symmetry. 
This relationship between $K_{s,ab}$(300 K) and $p$ gives us an opportunity to determine $p$ separately for OP and IP in multilayered cuprates, if it is confirmed that the relationship holds in multilayered cuprates.

\subsection{Hole density estimation in multilayered compounds}
\begin{table*}[htbp]
\caption[]{List of $K_{s,ab}$(300 K), $p'$, and $B$ for multilayered compounds: three-layered 0223F, Hg1223 \cite{MagishiHg1223}, and five-layered Hg1245 \cite{Kotegawa2004}. Here, $p'$ is the hole density that is tentatively evaluated by using experimental values of $K_{s,ab}$(300 K) and the solid line in Fig.~\ref{fig:Bterm}(a).}
\label{samples}
\begin{center}
{\tabcolsep = 3.5mm
  \renewcommand\arraystretch{1.2}
  \begin{tabular}{ccccc}
    \hline\hline
Sample, $T_c$   &layer   &   $K_{s,ab}$(300 K) ($\%$) &   $p'$      &  $B$(kOe/$\mu_B$)     \\
    \hline
0223F, 120 K    &   OP   &   0.35                     &     0.156   &   72          \\
                 &   IP   &   0.28                     &     0.115   &   67          \\
Hg1223, 133 K   &   OP   &   0.41                     &     0.185   &   90          \\
                 &   IP   &   0.32                     &     0.139   &   80          \\
Hg1245, 108 K   &   OP   &   0.34                     &     0.151   &   74          \\
                 &   IP   &   0.21                     &     0.075   &   61          \\
    \hline\hline
    \end{tabular}}
 \end{center}
\label{t:ggg2}
\end{table*}

In multilayered cuprates, doped hole carriers reside on OP and IP with different doping levels, and the CuO$_2$ layers show different physical properties due to the charge distribution even in the same sample. Therefore, it is required to separately estimate $p$ for OP and IP in order to study the electronic states of multilayered cuprates. 
It is invalid to apply the estimation methods used in single- and bi-layered compounds, for example, the relationship $T_c$=$T_{c,max}$[1-82.6($p$-0.16)$^2$] \cite{Presland}, the thermoelectric power \cite{Obertelli,Tallon}, and the bond valence sums \cite{Brown,Cava,TallonBV}. Those methods are applicable to evaluate total hole density, but not to evaluate each hole density at CuO$_2$ planes.
On the other hand, Cu-NMR measures the respective values of $K_{s,ab}$(300 K) for OP and IP. If the $K_{s,ab}$(300 K)-$p$ relationship in Fig.~\ref{fig:Bterm}(a) holds even in multilayered cuprates, it allows us to separately estimate $p$ for OP and IP from experimental values of $K_{s,ab}$(300 K).

Table \ref{t:ggg2} lists $K_{s,ab}$(300 K), $p'$, and $B$ for multilayered cuprates: three-layered 0223F, HgBa$_2$Ca$_2$Cu$_3$O$_{8+\delta}$ (Hg1223) \cite{MagishiHg1223}, and five-layered HgBa$_2$Ca$_2$Cu$_3$O$_{12+\delta}$ (Hg1245) \cite{Kotegawa2004}. Here, $p'$ is the hole density tentatively estimated by using the $K_{s,ab}$(300 K)-$p$ relationship in Fig.~\ref{fig:Bterm}(a). We plot $B$ of those compounds as crosses ($\times$) in Fig.~\ref{fig:Bterm}(b). The data plots fit into the other data, suggesting that the $K_{s,ab}$(300 K)-$p$ relationship is also valid in multilayered cuprates.
We may consider the case that the $K_{s,ab}$(300 K)-$p$ relationship is modified in multilayered cuprates. According to Eq.~(\ref{eq:Ks}), the possible source of the modification is the $p$ dependence of $B$ or of $\chi_s$ or both.
Experimental $B$ values in multi-layered compounds, however, fall on the same universal curve as shown in Fig.~\ref{fig:Bterm}(b); some unexpected modifications on the side of $\chi_s$ can indeed be considered as unlikely.
Therefore, we conclude that the $K_{s,ab}$(300 K)-$p$ relationship in Fig.~\ref{fig:Bterm}(a) is valid for CuO$_2$ planes in multilayered compounds, regardless of OP and IP.

Moreover, when we apply the $K_{s,ab}$(300 K)-$p$ relationship to five-layered compounds, 
the fact that the maximum $T_c$ takes place at $p$ $\sim$ 0.16 \cite{Mukuda2008} supports the validity of the application to multilayered compounds. 
Several attempts have been used to determine $p$ of hole-doped cuprates so far -- the relationship between $p$ and $K_{s,ab}$(300 K) in this work is a promising approach to estimate $p$, which is {\it applicable to multilayered compounds}. To our knowledge, it is the only method to separately determine $p$ for OP and IP in multilayered compounds with a reliable accuracy.

\section{Conclusion}

We have shown that the planar CuO$_2$ hole densities $p$ in high-$T_c$ cuprates are consistently determined with the Cu-NMR Knight shift. It has been demonstrated that the spin part of the Knight shift $K_{s,ab}$(300 K) at room temperature is material-independent in CuO$_2$ planes with tetragonal symmetry, and that $K_{s,ab}$(300 K) monotonically increases with $p$ from underdoped to overdoped regions. These observations suggests that $K_{s,ab}$(300 K) is a reliable method for determining planar CuO$_2$ hole densities.
The experimental values of $K_{s,ab}$(300 K) and of hyperfine magnetic fields for three-layered Ba$_2$Ca$_2$Cu$_3$O$_6$(F,O)$_2$ and other multilayered compounds support the application of the $p$-$K_s$(300 K) relationship to multilayered compounds. We remark that, to our knowledge, the relationship is the only method to separately determine $p$ for OP and IP in multilayered compounds with a reliable accuracy.

\section*{Acknowledgement}

The authors are grateful to M. Mori for his helpful discussions. This work was supported by Grant-in-Aid for Specially promoted Research (20001004) and by the Global COE Program (Core Research and Engineering of Advanced Materials-Interdisciplinary Education Center for Materials Science) from the Ministry of Education, Culture, Sports, Science and Technology (MEXT), Japan.


\clearpage

\end{document}